\documentclass[prb,preprint]{revtex4-1}

\usepackage{bbold}
\usepackage{graphicx}
\usepackage{bm}
\usepackage{amsmath}
\usepackage{color}
\usepackage{soul}

\newcommand{\vac}[1]{{\bf{#1}}}
\newcommand{\be}{\begin{equation}}
\newcommand{\ee}{\end{equation}}
\newcommand{\bey}{\begin{eqnarray}}
\newcommand{\eey}{\end{eqnarray}}

\newcommand{\bpi}{\boldsymbol{\pi}}
\newcommand{\bPi}{\boldsymbol{\Pi}}
\newcommand{\blambda}{\boldsymbol{\lambda}}
\newcommand{\bLambda}{\boldsymbol{\Lambda}}
\newcommand{\bkappa}{\boldsymbol{\kappa}}
\newcommand{\bnabla}{\boldsymbol{\nabla}}

\begin{document}

\title{{Gauge transformations and conserved quantities in classical and quantum mechanics}}

\author{Bertrand Berche}
\email{bertrand.berche@univ-lorraine.fr}
\affiliation{Groupe de Physique Statistique, Institut Jean Lamour, Universit\'e de Lorraine, 54506 Vandoeuvre-les-Nancy, France}
\affiliation{Centro de F\'isica, Instituto Venezolano de Investigaciones Cient\'ificas, 21827, Caracas, 1020 A, Venezuela}

\author{Daniel Malterre}
\email{daniel.malterre@univ-lorraine.fr}
\affiliation{Groupe Surfaces et spectroscopies, Institut Jean Lamour, Universit\'e de Lorraine, 54506 Vandoeuvre-les-Nancy, France}

\author{Ernesto Medina}
\email{ernestomed@gmail.com}
\affiliation{Centro de F\'isica, Instituto Venezolano de Investigaciones Cient\'ificas, 21827, Caracas, 1020 A, Venezuela}
\affiliation{Groupe de Physique Statistique, Institut Jean Lamour, Universit\'e de Lorraine, 54506 Vandoeuvre-les-Nancy, France}

\date{\today}

\begin{abstract}
We are taught that gauge transformations in classical and quantum mechanics do not change the physics of the problem.  Nevertheless here we
discuss three broad scenarios where under gauge transformations: (i) conservation laws are not preserved in the usual manner; (ii) non-gauge-invariant
quantities can be associated with physical observables; and (iii) there are changes in the physical boundary conditions of the wave function
that render it non-single-valued.  We give worked examples that illustrate these points, in contrast to general opinions from classic texts.  We also give a
historical perspective on the development of Abelian gauge theory in relation to our particular points. Our aim is to provide a discussion of these issues at the graduate level. 
\end{abstract}



\maketitle

\section{Introduction\label{sec1}}

It is hard to exaggerate the role of gauge invariance in the construction of physical theories, and many aspects of gauge theory, gauge co-variance, gauge invariance, and the connection to symmetry and conservation laws have been discussed both in textbooks and research journals.
The aim of this paper is to attempt to clarify some subtleties that arise in quantum mechanics in the context of gauge transformations: Is the wave function always single-valued? If not, what are the consequences of its multi-valued character on the definition of observables? 
We also discuss the physical meaning of some gauge-invariant
and non-gauge-invariant quantities in connection with rotational symmetry. This paper is intended to be followed by graduate students and may serve as the basis for advanced exercises or projects in a second course on nonrelativistic quantum mechanics.

Gauge invariance was originally discovered as a property of Maxwell's equations in electrodynamics, where the equations of the theory do not change when a gauge transformation of the potentials is performed: $\vac A\to\vac A'=\vac A+\bnabla\alpha$, $\phi\to\phi'=\phi-\partial_t\alpha$, where $\mathbf{A}$ is the vector potential, $\phi$ is the scalar potential, and $\alpha$ is an arbitrary function of the space and time coordinates.  According to a widespread teaching paradigm, this freedom appears mostly as a device that can help simplify a problem mathematically while leaving the physical content intact, i.e., with the same electric and magnetic fields. This view probably goes back to the work of Heaviside:\cite{WuYang06} 

\begin{quote}
$\vac A$ and its scalar potential parasite $\phi$ sometimes causing great mathematical complexity
and indistinctiveness; and it is, for practical reasons, best to murder the
whole lot, or, at any rate, merely employ them as subsidiary functions$\,\ldots$
\end{quote} 

This opinion was nevertheless not held by Maxwell or Thomson, who considered $\vac A$ to be a momentum per charge (i.e., more than a \textit{subsidiary} function), and there has been an abundant literature, in particular in this journal,\cite{Calkin,Konopinski,SemonTaylor,Johnson} discussing the role of $e\vac A$ as a linear momentum in a similar manner to $e\phi$ as a potential energy.  

With the advent of quantum theory, the role of the vector potential was intensely revisited,\cite{O'Raifeartaigh} in particular with the celebrated paper of Aharonov and Bohm.\cite{AharonovBohm}   An account of the most relevant literature is given in the Resource Letter of Cheng and Li.\cite{ChengLi}

An important new insight regarding gauge theory was achieved by Weyl in 1918, and then in 1929, when he considered a generalization of the gravitation theory of Einstein.~\cite{Weyl1918} 
While lengths of vectors are conserved in Riemannian geometry, Weyl allowed for a length change during parallel transport and thus introduced an additional connection, which he proposed to identify with the electromagnetic gauge vector, providing the first unified theory of gravitation and electromagnetism. This theory did not survive major physical objections at the time,\cite{Pauli1958} but became prominent after its reformulation in the context of quantum mechanics.\cite{Weyl1929} There, it is the wave function that inherits a phase in an electromagnetic field, suggesting the possibility of reformulating Weyl's theory by contemplating complex objects instead of vectors in Riemann space. This new concept gave birth to modern gauge field theory.\cite{Weyl1929}

Before becoming a standard approach in textbooks,\cite{YangMills54,Ryder,Ramond}
Weyl's theory was spread in the physics community through influential papers by Dirac\cite{Dirac31} and Pauli,\cite{Pauli} and then by Wu and Yang.\cite{WuYang75} In the spirit of Einstein's theory of gravitation, it converts an interaction into a property of the ``space,'' in other words, it ``geometrizes'' the electromagnetic interaction via the so called non-integrable phase of the wave function.  Non-integrability here means non-definite values for the phase at points on a space-time trajectory.  Only changes in the phase (and thus its derivatives) have meaning.  The derivatives of the phase are in fact the gauge fields in electromagnetism. 

If the phase of the wave function is non-integrable, an issue arises concerning the single-valuedness (or multi-valuedness) of wave functions in geometries where it closes on itself---a question that is usually overlooked in the literature.  Even influential textbooks have contradictory statements concerning this delicate question.  Some authors consider the single-valuedness of the wave function as mandatory:

\begin{quote}The conditions which must be satisfied by solutions of Schr\"odinger's equation are very general in character. First of all, the wave function must be single-valued and continuous in all space.\cite{Landau}\end{quote}

\begin{quote}It is implicit in the fundamental postulates of quantum mechanics that the wave function for a particle without spin must have a definite value at each point in space. Hence, we demand that the wave function be a \textit{single-valued} function of the particle's position.\cite{Merzbacher}\end{quote}

The condition of single-valuedness is often considered as a prerequisite to build eigenstates of angular momentum, leading to integer eigenvalues of $L_z$ (in units of $\hbar$).\cite{Cohen,Messiah,Liboff,MerzbacherAJP}

Other authors consider this question with more caution, for example:

\begin{quote}It is reasonable to require that the wave function and its gradient be continuous, finite, and single-valued at every point in space, in order that a definite physical situation can be represented uniquely by a wave function.~\cite{Schiff}\end{quote}

\begin{quote}Because kets (or wave functions) are not in themselves observable quantities,
they need not be single-valued. On the other hand, a Hermitian operator $A$ that
purports to be an observable must be single-valued under rotation to insure that its
expectation value $\langle\psi|A|\psi\rangle$ is single-valued in an arbitrary state.~\cite{Gottfried}\end{quote}

\begin{quote}Multiple-valued wave functions cannot be excluded a priori. Only physically
measurable quantities, such as probability densities and expectation values of
operators, must be single-valued. Double-valued wave functions are used in the
theory of particles with intrinsic spin.~\cite{Weisskopf}\end{quote}

Ballentine addresses the question without hiding the underlying difficulties:
 
\begin{quote}The assumptions of \textit{single-valuedness} and \textit{nonsingularity} can be justified in a classical field theory, such as electromagnetism, in which the field is an observable physical quantity. But in quantum theory, the state function $\Psi$ does not have such direct physical significance, and the classical boundary conditions cannot be so readily justified. \textit{Why should $\Psi$ be single-valued under rotation?} Physical significance is attached, not to $\Psi$ itself, but to quantities such as $\langle\Psi|A|\Psi\rangle$, and these will be unchanged by a $2\pi$ rotation\,\dots. \textit{Why should $\Psi$ be nonsingular?} It is clearly desirable for the integral of $|\Psi|^2$ to be integrable so that the total probability can be normalized to one\,\dots. It is difficult to give an adequate justification of the conventional boundary conditions in this quantum-mechanical setting.\cite{Ballentine}\end{quote}

In this paper we will first introduce the problem via a discussion of gauge invariance in the classical context.  We will then briefly review the extension to quantum mechanics, involving both the operators and the wave function, and define gauge-invariant and non-gauge-invariant
quantities and state conservation laws. This discussion will set a precise stage to illustrate both changes in the statement of conservation
laws and lack of single-valued wave functions under certain gauge transformations.

\section{Gauge invariance, gauge covariance, and unitary transformations}\label{1bis}

Consider a single nonrelativistic spinless particle in an external magnetic field $\vac B$. For simplicity, we ignore the scalar potential in this discussion. The corresponding classical Hamiltonian is
\be H=\frac{1}{2m}(\vac p-e\vac A(\vac r))^2,\ee
where $(\vac r,\vac p)$  are the fundamental dynamical variables in the Hamiltonian formulation, i.e., the position $\vac r$ and the \textit{canonical momentum} conjugate to the position, $\vac p=\partial L/\partial\dot {\vac r}$ with $L$ the Lagrangian of the particle.
Newtonian mechanics dictates that the physical quantities experimentalists can measure are the
positions and velocities $\vac r$ and $\vac v$, and one can define a \textit{mechanical momentum} as ${\vac \bpi}=\vac p-e\vac A=m{\vac v}$ in terms of which the Hamiltonian reduces to purely kinetic energy,
$H=\bpi^2/2m$.
If one changes the gauge that determines the potentials in the Hamiltonian according to $\vac A\to\vac A'=\vac A+\bnabla\alpha$, with $\alpha$ a function depending on space (and time in the more general case),
the invariance of the physics is stated as
\begin{eqnarray}
\vac r'&=&\vac r, \label{gaugeclassicalr}\\
\vac \bpi'&=&\vac \bpi,
\label{gaugeclassical}
\end{eqnarray}
where a prime denotes the physical quantity in the new gauge.
This condition entails a gauge dependence in the canonical momentum,
\be\vac p'=\vac p+e\bnabla \alpha.\label{gaugeclassicalp}
\ee
As can be  seen, $\vac \bpi$ does not change with the gauge choice because the change in $\vac p$ is compensated by the change in $\vac A$. All gauge-invariant physical quantities are thus built from functions of $\vac r$ and $\vac \bpi$.  All the classical physical quantities are then specified by  combinations of these mechanical variables. A familiar example of a function one can build is the angular momentum. The canonical function would be built as $\vac l=\vac r\times\vac p$ while the mechanical angular momentum would be $\blambda=\vac r\times m\vac v=\vac r\times \bpi$. The latter is gauge-invariant by construction, while the former, like $\vac p$, is not.

In quantum mechanics the quantization rules dictate that we now make the replacements $\vac r\rightarrow  {\hat{\vac R}}$ and
$\vac p\rightarrow  {\hat{\vac P}}$ of dynamical variables with the corresponding operators  {(denoted with ``hats'')}.  From these
dynamical quantized variables one can then build the gauge potential $\vac A( {\hat{\vac R}})$, the velocity operator
$\vac V( {\hat{\vac R}},\hat{\vac P})=(\hat{\vac P}-e\vac A)/m$, and also the mechanical momentum operator $\hat{\bPi}=m\hat{\vac V}$,
as well as $ {\hat{\ \vac L}=\hat{\vac R}\times\hat{\vac P}}$ and $ {\hat{\bLambda}=\hat{\vac R}\times \hat{\bPi}}$ for the corresponding canonical and mechanical angular momenta. We then have
the quantized Hamiltonian
\begin{equation}
 {\hat H}=\frac{1}{2m}( {\hat{\vac P}}-e\vac A( {\hat{\vac R}}))^2.
\end{equation}
The  canonical operators obey commutation rules  {$[X_j,P_k]=i\hbar\delta_{jk}$, $j,k=1,2,3$, where $X_j$ and $P_k$ are the Cartesian components of ${\hat{\vac R}}$ and $ {\hat{\vac P}}$,} and, as a conventional rule, it is convenient to preserve the
form of the canonical momentum operator in the position representation $\hat{\vac P}=-i\hbar\bnabla$ for all gauge choices. This is also a consequence of the fact that the canonical momentum $\hat{\vac P}$ is the generator of space translations, and this property should be kept for all gauges. So the counterparts of Eqs.\ (\ref{gaugeclassicalr}) and (\ref{gaugeclassical}) are
\begin{eqnarray}
\hat{\vac R}'&=&\hat{\vac R},\label{gaugequantumR} \\
\hat{\vac P}'&=&\hat{\vac P},
\label{gaugequantumP}
\end{eqnarray}
and they entail that 
\be \hat{\bPi}'=\hat{\bPi}-e\bnabla\alpha.\label{gaugePi}\ee
Now in quantum mechanics, the physical information is not only in the (operators representing) dynamical variables themselves, but also
in the expectation values, which involve the wave functions. In terms of the expectation values, the
rules of Eqs.\ (\ref{gaugequantumR}) and (\ref{gaugequantumP}) turn into the classical gauge results (see Eqs. (\ref{gaugeclassicalr})-(\ref{gaugeclassicalp})),
\begin{eqnarray}
\langle\psi'|\hat{\vac R}'|\psi'\rangle&=&\langle\psi|\hat{\vac R}|\psi\rangle, \label{gaugeexpectationvalueR}\\
\langle\psi'|\hat{\vac P}'|\psi'\rangle&=&\langle\psi|\hat{\vac P}+e\bnabla\alpha|\psi\rangle. \label{gaugeexpectationvalueP}
\end{eqnarray}

One can arrive at the same conclusion by cooking up the appropriate unitary transformation $\hat U$
designed so that
\begin{equation}
\psi'({\vac R})=\hat U\psi({\vac R}),\label{TransfPsi}
\end{equation} 
with $\hat U\hat U^{\dagger} {=\hat U^{\dagger}\hat U}=1$ to preserve the norm of $\psi$.
If we are to satisfy Eqs.\ (\ref{gaugeexpectationvalueR}) and (\ref{gaugeexpectationvalueP}) then we must have
\begin{eqnarray}
\hat U^{\dagger}\hat{\vac R} \hat U&=&\hat{\vac R},
\\
\hat U^{\dagger}\hat{\vac P} \hat U&=&\hat{\vac P}+e\bnabla\alpha.
\end{eqnarray}
These two equations are satisfied by the choice~\cite{Cohen321}
\be \hat U=\exp\left({i\frac{e}{\hbar}\alpha}\right),\ee
 where we again stress that  $\alpha$ depends on $\vac r$ and would in the general case also depend on $t$.

In the case of the usual dynamical variables $\hat{\vac R}$ and $\hat{\bPi}$, 
one has
\begin{eqnarray}
\hat U\hat{\vac R}\hat U^{\dagger}&=&\hat{\vac R}\hat U^{\dagger}\hat U=\hat{\vac R},\label{unitaryOfR}\\
\hat U\hat{\vac \bPi}\hat U^{\dagger}&=&\hat U\vac (\hat{\vac P}-e\vac A(\hat{\vac R}))\hat U^{\dagger}
=\hat{\vac \bPi} -e\bnabla \alpha,
\label{operatorunitary}
\end{eqnarray}
since $\hat U$ is only a function of the position operator. These relations coincide with the gauge-transformed counterparts given in Eqs.~(\ref{gaugequantumR}) and (\ref{gaugePi}).  This is an important property, which has to do with the gauge invariance of position and mechanical momentum, as we now discuss.

Most of the physical quantities ${\cal Q}$ in the theory (here we omit the spin and any other internal properties of the particle) can be expressed in terms of the fundamental dynamical variables, which, in Hamiltonian formalism, are $\hat{\vac R}$ and $\hat{\vac P}$, i.e., ${\cal Q}(\hat{\vac R},\hat{\vac P})$. They are represented by  Hermitian operators $\hat Q$, usually referred to as  \textit{observables}.  (See, for example, Ref.~\onlinecite{Messiah2} for a discussion of the general relation between observables and Hermitian operators). Under gauge transformations, they become $\hat{Q}'\equiv {\cal Q}(\hat{\vac R}',\hat{\vac P}')$.
Let us consider such an operator $\hat Q$ that has the additional property of being \textit{gauge invariant}, that is, its matrix elements, being possibly associated with the results of measurements, do not depend on the gauge choice: 
\be \langle\psi'|\hat{Q}'|\psi'\rangle=\langle\psi |\hat Q|\psi\rangle.\label{eq-gaugeinvmatel}\ee This requires  $\hat Q=\hat U^\dagger\hat Q'\hat U$, or 
\be \hat Q'=\hat U\hat Q\hat U^\dagger.\label{eq-Q}\ee  
This relation is fundamental to understanding gauge invariance in quantum mechanics. The observable ${\cal Q}$ is gauge \textit{invariant} in the sense that any matrix element takes the same value in different gauges~(\ref{eq-gaugeinvmatel}), but the operator $\hat Q$ representing the quantity has then to be gauge \textit{covariant} (\ref{eq-Q}) in order to achieve this property.  This requirement can be satisfied by operators that keep the same form in different gauges, e.g., $\hat{\vac R}$ in Eq.~(\ref{unitaryOfR}), as in the classical realm.  But it can also be satisfied by operators that differ in the two gauges, unlike the classical case (e.g., $\hat{\bPi}$  in Eq.~(\ref{operatorunitary})). On the other hand, there also exist operators that keep the same form in two gauges, but that do not obey the gauge covariance property (\ref{eq-Q}) and hence are not gauge invariant (e.g., $\hat{\vac P}'$ in Eq.~(\ref{gaugequantumP}) does not coincide with $\hat U\hat{\vac P}\hat U^\dagger=\hat{\vac P}-e\bnabla\alpha$).
In Table~I we list different physical properties that are modified (or not) by gauge transformations.

Some authors consider  gauge-invariant quantities as ``genuine'' physical quantities, and consider   non-gauge-invariant quantities to be not genuinely physical.\cite{Cohen321} 
However, some non-gauge-invariant Hermitian operators, like the canonical linear momentum or canonical angular momentum operators, play fundamental roles in physics. They are the generators of the space group (infinitesimal translation and rotation operators, respectively), and thus, as  conserved quantities in closed systems, are central in the Hamiltonian formalism. 
Moreover, according to Noether's theorem,\cite{BjorkenDrell,Sakurai,WeinbergQM} such quantities are conserved in physical situations where the corresponding symmetry is satisfied by the Hamiltonian.
Because conservation laws are of critical importance, we adopt a less extreme position by stating that both gauge-invariant and non-gauge-invariant observables can be associated with physical quantities.  We will see below an example of a non-gauge-invariant operator that has a physical interpretation.

\begin{table}[h!]
\centering
\caption{Summary of gauge transformations of essential physical quantities. Note that the last relations for the gauge transformation of the Hamiltonian are written here in the general case of a space- and time-dependent gauge transformation (see Eq.~(\ref{EqHGal})).}
\begin{ruledtabular}
 \begin{tabular}{ l  cc l  l }
  Classical context &
 &&
   	\multicolumn{2}{c}{\hskip-4em Quantum context }\\
     \hline
      \multicolumn{5}{l}{Gauge vector:}\\
      $\vac A'=\vac A+\bnabla\alpha$ && && 
      $\vac A'=\vac A+\bnabla\alpha$\\
\hline
   \multicolumn{5}{l}{Gauge-invariant quantities:}\\
   
$\vac r'=\vac r$   &
&\qquad\qquad&
  $\langle\psi'|\hat{\vac R}'|\psi'\rangle=\langle\psi|\hat{\vac R}|\psi\rangle$ &
 $\hat{\vac R}'=\hat U\hat{\vac R}\hat U^{\dagger}=\hat{\vac R}$ \\

$\bpi'=\bpi$   &
&& 
$\langle\psi'|\hat{\bPi}'|\psi'\rangle=\langle\psi|\hat{\bPi}|\psi\rangle$&
 $\hat{\bPi}'=\hat U\hat{\bPi}\hat U^{\dagger}=\hat{\bPi}-e\bnabla\alpha$ \\

$\blambda'=\blambda$   &
&& 
$\langle\psi'|\hat{\bLambda}'|\psi'\rangle=\langle\psi|\hat{\bLambda}|\psi\rangle$ &
 $\hat{\bLambda}'=\hat U\hat{\bLambda}\hat U^{\dagger}=\hat{\bLambda}-e\hat{\vac R}\times\bnabla\alpha$ \\

$K'=K$   &
&& 
$\langle\psi'|\hat{K}'|\psi'\rangle=\langle\psi|\hat{K}|\psi\rangle$ &
 $\hat{K}'=\hat U\hat{K}\hat U^{\dagger}=\frac {1}{2m}(\hat{\bPi}-e\bnabla\alpha)^2$ \\

   \hline
      \multicolumn{5}{l}{Generators of space-time symmetries (regular gauge transformations):}\\

      $\vac p'=\vac p+e\bnabla\alpha$   &
      && 
      $\langle\psi'|\hat{\vac P}'|\psi'\rangle=\langle\psi|\hat{\vac P}+e\bnabla\alpha|\psi\rangle$ &
$\hat{\vac P}'=\hat{\vac P}=-i\hbar\bnabla\not=\hat U\hat{\vac P}\hat U^{\dagger}$ \\

      $\vac l'=\vac l+e\vac r\times\bnabla\alpha$   &
      && 
      $\langle\psi'|\hat{\vac L}'|\psi'\rangle=\langle\psi|\hat{\vac L}+e\hat{\vac R}\times\bnabla\alpha|\psi\rangle$ \qquad\qquad &
$\hat{\vac L}'=\hat{\vac L}=-i\hbar\hat{\vac R}\times\bnabla\not=\hat U\hat{\vac L}\hat U^{\dagger}$ \\

$H'=H-e\partial_t\alpha$ &
&&  $\langle\psi'|\hat{H}'|\psi'\rangle=\langle\psi|\hat{H}-e\partial_t\alpha|\psi\rangle$ &
$\hat{H}'=\hat{H}=i\hbar\partial_t\not=\hat U\hat{H}\hat U^{\dagger}$\\

\end{tabular}
\end{ruledtabular}
\label{table}
\end{table}

Let us now discuss the effect of a gauge transformation on a conserved quantity. Assume that for some reason, a physical quantity ${\cal Q}$ should be conserved in a given problem, a property that one expresses in quantum mechanics by the equation
\be
\frac {d}{dt}\langle\hat Q\rangle_{\psi}=0,
\ee
where $\langle\hat Q\rangle_{\psi}$ is the expectation value of $\hat Q$ in the quantum state $|\psi\rangle$, that is, $\langle\hat Q\rangle_{\psi}=\langle\psi|\hat Q|\psi\rangle$. This implies
\be
\frac i\hbar\langle[\hat H,\hat Q]\rangle_\psi+\langle\partial_t\hat Q\rangle_\psi=0,
\ee
and, if $\partial_t\hat Q=0$, we have $\langle[\hat H,\hat Q]\rangle_\psi=0$. If we are furthermore working in a gauge 
such that $\hat Q$ commutes with $\hat H$, the equation is automatically fulfilled. The commutation of an observable with the Hamiltonian
implies then the conservation of that observable. As a consequence, the operators $\hat Q$ and $\hat H$ have in this case the same eigenstates.  

The conservation property should obviously be robust to gauge transformations. Hence in a different time-independent gauge 
with $|\psi'\rangle=\hat U|\psi\rangle$ and
\be 
\hat H'=\hat U\hat H\hat U^\dagger-e\partial_t \alpha= \hat U\hat H\hat U^\dagger,\label{EqHGal}
\ee 
it is straightforward to show that a gauge-invariant quantity obeying Eq.~(\ref{eq-Q}) commutes with $\hat H'$:
\be 
[\hat H',\hat Q']=[\hat U\hat H\hat U^\dagger,\hat U\hat Q\hat U^\dagger]=\hat U[\hat H,\hat Q]\hat U^\dagger=0.
\ee 
The case of a non-gauge-invariant conserved quantity is more subtle. Consider a quantity like $\hat{\vac P}$ or $\hat{\vac L}$ that satisfies 
\be 
\hat Q=\hat Q'\ne \hat U\hat Q\hat U^\dagger.
\ee 
It might appear that $[\hat H',\hat Q']\not=0$, i.e., $\hat Q'$ and $\hat H'$ do not share the same eigenstates.  Nevertheless $\hat Q'$ is still a conserved quantity in the sense that the {\em expectation value} of the commutator vanishes:
\be 
\frac {d}{dt}\langle \hat Q'\rangle_{\psi'}=\langle[\hat H',\hat Q']\rangle_{\psi'}=0.
\ee 

Let us illustrate this property, anticipating the example of angular momentum in a magnetic field with cylindrical symmetry along the $z$ direction, treated classically in Sec.~\ref{SecClass}, and quantum mechanically in Sec.~\ref{SecQM}.  Among all gauge choices, we can select one exhibiting the cylindrical symmetry of the problem so that $\hat L_z$ is a conserved quantity ($[\hat H,\hat L_z]=0$). For a different gauge choice obtained from the unitary operator $\hat U$, the corresponding conserved quantity is $\hat U\hat L_z\hat U^{\dagger}$ whereas the canonical angular momentum in this gauge is $\hat L'_z=\hat L_z$ and does not commute with $\hat H'$.  However as $\hat L'_z=\hat U\hat L_z\hat U^{\dagger}-e\partial_{\varphi}\alpha$ (cylindrical coordinates), it is straigthforward to show from the periodicity of $\alpha$ that $\langle[\hat H',\hat L_z']\rangle_{\psi'}=0$.
This discussion opens a new question since  there appears a particular gauge in which the conservation equation takes a simpler form, $[\hat H,\hat L_z]=0$ rather than $\langle[\hat H',\hat L_z]\rangle_{\psi'}=0$.  {This particular} gauge respects the space-time symmetry encoded in the conserved quantity, as we will illustrate in Sec.~\ref{SecQM}.

We thus have to distinguish two kinds of physical quantities, both corresponding to observables in quantum mechanics and represented by Hermitian operators.  The first ones are gauge-invariant and satisfy Eq.~(\ref{eq-Q}). They are associated with the same quantity in different gauges (like position and velocity) and can be simply measured and interpreted. The second ones, like the canonical momentum or canonical angular momentum,  are not gauge-invariant. They represent quantities that, being the generators of space-time transformations, keep the same geometrical meaning but carry different physical content in different gauges. Nevertheless they might be related to fundamental symmetries and then  {commute with the Hamiltonian} in the gauge where the Hamiltonian exhibits the total symmetry of the system.  We emphasize that the Hamiltonian itself is such a quantity, $\hat H=\hat K+e\phi$ (with $\hat K$ the kinetic energy).  {Although} not gauge-invariant in the general case  {due to the presence of the scalar potential contribution,}  {it governs the time evolution of the system} and its role in the physical theory can hardly be overestimated.

\section{A case study:  classical treatment}\label{SecClass}

We now consider the classical problem of a particle with charge $e$ subject to a central force and moving in a circular orbit of radius $\rho$.  In terms of unit vectors $\vac u_\rho$ and $\vac u_\varphi$, the particle's position is $\vac r=\rho\vac u_\rho$ and its velocity is $\vac v=v_\varphi\vac u_\varphi$.  We then turn on a uniform magnetic field perpendicular to  the plane of the trajectory.  We use a superscript~0 to denote the values of quantities before the application of the field, e.g., the radius $\rho_0$ and velocity $v_{0\varphi}$ are linked to the central force field $F$ by Newton's law, $F=mv_{0\varphi}^2/\rho_0$. Due to the applied magnetic field (approximated as a  time-dependent uniform field $\vac B=B(t)\vac u_z$),  a time-dependent gauge vector in the cylindrical gauge \be\vac A=\frac 12B(t)\rho\vac u_\varphi,\ee is the source of the electromotive force  {$e\vac E=-e\partial_t\vac A$ on the charge. This force is due to the action of the induced electric field $\vac E$ that arises because of a changing flux within the circular motion. }

If we consider the change in the gauge vector $\delta\vac A=\partial_t\vac A\ \!dt$ associated with the application in the time $dt$ of an infinitesimal magnetic field $\delta\vac B$, the electromagnetic force $-e\partial_t\vac A$
leads to a variation of kinetic energy $\delta(\frac 12 mv_\varphi^2)=mv_\varphi\delta v_\varphi=-e\delta\vac A\cdot\vac v$, hence a modification of the velocity $\delta v_\varphi=-(e/m)\delta\vac A\cdot\vac v/|\vac v|$, which depends on the relative orientation of $\vac v$ and~$\vac A$.
This variation of velocity due to the external field is exactly what is needed to keep the trajectory unchanged, because now the total force exerted on the charge is $F+ev_\varphi\delta B$ and coincides to first order with $m(v_{0\varphi}+\delta v_\varphi)^2/\rho$, with the same radius $\rho=\rho_0$ as in the initial state.
The action of the field modifies the charge's speed along the circular trajectory, hence causing a  change in the magnetic moment of the loop  that is \textit{opposite} to~$\vac B$.  This is the origin of orbital diamagnetism in this classical model.

This result is consistent with the conservation of the canonical angular momentum. The problem, as it was stated here, exhibits  rotational symmetry around the $z$ axis at any time and this implies that the canonical angular momentum $l_z=(\vac r\times\vac p)_z$ is conserved. Before the application of the field it is $l_{0z}=mv_{0\varphi}\rho_0$, while in the final state it is computed as $l_z=(mv_\varphi+eA_\varphi)\rho$. As $v_\varphi-v_{0\varphi}=-eA_\varphi/m$, conservation of canonical angular momentum $l_z=l_{0z}$ is ensured.
When the magnetic field is applied, the induced electric field, $-e\partial_t\vac A$, leads to a change of kinetic energy and of mechanical angular momentum. In the cylindrical gauge, the canonical angular momentum is conserved. But this is not a gauge-independent quantity and in another gauge that does not exhibit the symmetry of the problem, the canonical angular momentum would not be conserved. 

It is thus instructive to analyze the same problem with a different choice of gauge. Consider now the Landau gauge $\vac A'=B(t)x\ \!\vac u_y$. In cylindrical coordinates it is 
\be\vac A'=\frac 12B(t)\rho\sin(2\varphi)\ \!\vac u_\rho+\frac 12 B(t)\rho(1+\cos(2\varphi))\ \!\vac u_\varphi,\ee and we pass from $\vac A$ to $\vac A'$ via the gauge transformation $\vac A'=\vac A+\bnabla\alpha$ with 
\be
\alpha(\rho,\varphi)=\frac 14B(t)\rho^2\sin(2\varphi).\label{eq-alpha}
\ee
The vector potential $\vac A'$ is {\em not} uniform along the trajectory, and this breaks the rotational symmetry in  {the} formulation of the problem (e.g., the Hamiltonian explicitly depends on the angle~$\varphi$).  The Lagrangian of the particle,
\be L'=\frac 12m|\vac v|^2-e(\phi'-\vac v\cdot\vac A'),\ee
also depends explicitly on $\varphi$ and, as a consequence, the canonical angular momentum is not a conserved quantity.
 Although $-\partial_t\vac A'\not= -\partial_t\vac A$, the physical problem itself is nevertheless still the same, because in the new gauge there is an additional contribution to the electric field, $-\bnabla\phi'$, with $\phi'=-\partial_t\alpha$ in such a way that  the force exerted on the charge, $-e(\partial_t\vac A'+\bnabla\phi')$, is the same as $-e\partial_t\vac A$ in the cylindrical gauge.
The  {canonical} angular momentum {of the particle} in the new gauge can be calculated as $l'_z=(\vac r'\times(m\vac v'+e\vac A'))_z=mv_\varphi\rho+eA'_\varphi\rho=l_{0z}+\frac 12eB\rho^2\cos(2\varphi)$ (we use the fact that $\vac r$ and $\vac v$ are gauge-invariant), i.e., it is not conserved, and compared to its expression in the cylindrical gauge, one has
\be
l'_z=l_z+\frac{e}{2\pi}\Phi\cos(2\varphi),\label{eq-lzprime}
\ee
where $\Phi=B\pi\rho^2$ is the magnetic flux enclosed by the loop. 
In the non-rotationally-symmetric gauge, the canonical angular momentum $l'_z(t)$ oscillates around an average value that is its value $l_z$ in the cylindrical gauge. 
 
 There is another way to see what is happening between the two choices of gauge, following an interpretation given by Feynman.\cite{Feynman}
The full system under consideration is the particle \textit{and} the field. In the initial state, there is no field and the total canonical angular momentum reduces to the particle's mechanical contribution $mv_{0\varphi}\rho_0$. In the final state where the applied field $\vac B$ has reached its final static value,  the contributions to the mechanical momentum can be written for the particle as $mv_{\varphi}\rho$, and for the field~\cite{CohenAtoms} as 
\be \varepsilon_0\int\vac r'\times (\vac E_{e}(\vac r')\times \vac B(\vac r'))d^3r',\label{EqFieldContrib}\ee 
where  $\vac E_e$ is the Coulombic contribution of the particle of charge $e$.
Equation~(\ref{EqFieldContrib}) corresponds to the angular momentum transfer from the field to the particle via $\vac E_e(\vac r')$
 (note that in an intermediate state when $\vac B$ depends on time, the associated electric field would also contribute to the  field angular momentum).
 Due to the Coulombic form of 
 \be\vac E_e(\vac r')=\frac{e}{4\pi\varepsilon_0}\frac{\vac r'-\vac r}{|\vac r'-\vac r|^3},\ee  {the expression written in Eq.~(\ref{EqFieldContrib}) takes the form}~\cite{Johnson,Konopinski}
 \be
 \frac{e}{4\pi}\int \vac r'\times \left(\frac{\vac r'-\vac r}{|\vac r'-\vac r|^3}\times\vac B(\vac r') \right)d^3r'=\vac r\times e\vac A_{\rm sym.}(\vac r),
 \ee
with \be\vac A_{\rm sym.}(\vac r)=\frac{\mu_0}{4\pi}\int\frac{\vac j(\vac r')}{|\vac r-\vac r'|}d^3r',\ee the vector potential at the particle's position $\vac r$ {\rm in the cylindrical gauge}, i.e., with our notations $\vac A_{\rm sym.}(\vac r)=\vac A(\vac r)$.
The quantity that is conserved is the canonical angular momentum
\be \vac l=\vac r\times (\vac v+e\vac A_{\rm sym}(\vac r)).\label{Eqlcan}\ee

With another gauge choice $\vac A'$, obviously the particle's contribution to the mechanical angular momentum is unchanged, and similarly, the field's contribution (\ref{EqFieldContrib}) is also unchanged since it only depends on $\vac E$ and $\vac B$ fields, but now
\be 
\vac l'=\vac r\times (\vac v+e\vac A'(\vac r)),\label{Eqlprimecan}
\ee
differs from Eq.~(\ref{Eqlcan}). Note that subtracting Eq.~(\ref{Eqlcan}) from Eq.~(\ref{Eqlprimecan}), we recover Eq.~(\ref{eq-lzprime}) via the gradient of the gauge function $\alpha$ in Eq.~(\ref{eq-alpha}).

The fact that conservation of angular momentum is gauge dependent has been discussed in detail in the 
 literature.\cite{Konopinski,SemonTaylor} In the cylindrical gauge, the Hamiltonian has the full symmetry of the physical 
 problem (we apply  an axially symmetric magnetic field here). The solution of the problem exhibits the
 full symmetry and conservation of angular momentum is satisfied, i.e., $l_z(t)={\rm const}$. In the Landau gauge,
 the Hamiltonian (or Lagrangian) displays a lower symmetry, which manifests itself as a gauge-dependent oscillation that reflects the original conservation law only on average.
 
The vector potential acquires a physical significance in the symmetric gauge as the linear momentum transfer from the field to the charge.~\cite{Zangwill}
This example shows how to interpret physically a non-gauge-invariant quantity, the canonical angular momentum of the particle, or the vector potential, in a particular gauge.

\section{Quantum mechanical formulation}\label{SecQM}

\subsection{Ring and regular gauge transformations}\label{sec2}

Let us now illustrate the same concepts with the same problem treated in quantum mechanics.
Consider  a single electron without spin moving freely on a circular ring of radius $\rho=a$ in cylindrical coordinates $(\rho,\varphi,z)$. The eigenfunctions are standing waves on the ring,
$\psi_{0l_{z}}(\varphi)=(2\pi a)^{-1/2}e^{il_{z}\varphi}$, which are also eigenstates of the canonical angular momentum $\hat{L}_z=-i\hbar\partial_\varphi$ with eigenvalues $\hbar l_{z}$ with $\l_{z}\in\mathbb{Z}$.

The subscript $0$ is for the initial state of the system with no magnetic field.
In the presence of a uniform magnetic field $\vac B=B\ \!\vac u_z$ piercing the ring with a flux
$\Phi=B\pi a^2$, with the choice of the cylindrical gauge where the gauge vector takes the form 
$\vac A=\frac 12B\rho\vac \ \!\vac u_\varphi$,  the Hamiltonian on the ring in the nonrelativistic limit reduces to 
\bey \hat H&=&\frac{1}{2m}(\hat P_\varphi-eA_\varphi(a))^2\nonumber\\
&=&\frac{\hbar^2}{2ma^2}(-i\partial_\varphi-\Phi/\Phi_0)^2,\label{eqH}\eey 
with the ordinary representation of the canonical momentum 
\be \hat P_\varphi=-i\hbar a^{-1}\partial_\varphi,\label{eqp}\ee 
and $\Phi_0=2\pi\hbar/e$ the quantum unit of flux.
Thanks to rotational symmetry, $[\hat H,\hat L_z]=0$ (the Hamiltonian exhibits the symmetry of the physical problem), the eigenfunctions of $\hat H$ are again those of the canonical angular momentum $\hat L_z$,
namely 
\be\psi_{l_z}(\varphi)=(2\pi a)^{-1/2}e^{il_z\varphi},\label{eqpsi}\ee with integer $l_z$ values, and  the eigenenergies are 
\be E_{l_z}(\Phi)=\frac{\hbar^2}{2ma^2}(l_z-\Phi/\Phi_0)^2,
\ee
while the eigenvalues of the canonical angular momentum $\hbar l_z$ are unchanged (i.e., conserved).
The eigenfunctions are single-valued, 
 $\psi_{l_z}(\varphi+2\pi)=\psi_{l_z}(\varphi)$, i.e., they belong to the Hilbert space with specified boundary conditions
\be
{\cal H}=\{\psi(\varphi)
\ | \ \textstyle
\int_0^{2\pi}|\psi|^2\,d\varphi <+\infty \ \hbox{and}\  \psi(\varphi+2\pi)=\psi(\varphi)\}.\label{eqHilbert}\ee
The magnetic field, breaking time reversal symmetry, induces 
 {an electron current. 
 This contribution} to the  persistent current in the ring is given by the $\Phi$-dependent part of the corresponding energy eigenvalue,
\bey
j_\varphi&=&-\frac{\partial E_{l_z}}{\partial\Phi}\nonumber\\
&=&\frac e{m}{\psi_{l_z}}^*(\varphi)(-i\hbar a^{-1}\partial_\varphi-eA_\varphi(a))\psi_{l_z}(\varphi)\nonumber\\
&=&\frac{e\hbar}{2\pi ma^2}(l_z-\Phi/\Phi_0).
\eey
We note that, like the mechanical linear momentum, a \textit{mechanical angular momentum}, 
$\hat \Lambda_z=-i\hbar\partial_\varphi-eaA_\varphi(a)$, related to the angular velocity,  {appears in this equation.} It is the relevant quantity needed to calculate a physical quantity like the current density, but $\hat L_z$ is the operator associated with the conservation law in the cylindrical gauge.

\begin{figure}[ht]
\begin{center}
\includegraphics[width=8cm]{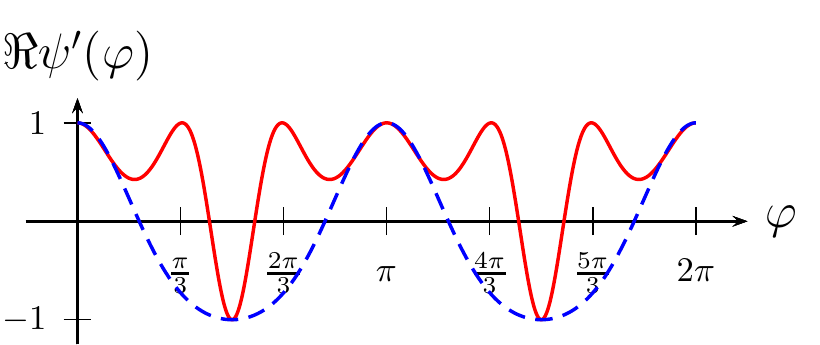}
\end{center}
\caption{The real part of the Landau-gauge $l_z=2$ eigenfunction, for $\Phi/\Phi_0=-2.5$ (solid) and $1/3$ (dashed). }\label{fig1}
\end{figure}

Instead of the cylindrical gauge $\vac A$, one can use the Landau gauge 
$\vac A'=
\frac 12 B\rho \sin (2\varphi)\ \!\vac u_\rho +\frac 12 B\rho(1+\cos (2\varphi))\ \!\vac u_\varphi$, even though the latter choice is again not adapted to the circular geometry. 
Equation~(\ref{eq-alpha}) is the gauge function that describes the change of formulation from $\vac A$ to $\vac A'$
and the eigenfunctions on the ring are modified accordingly, 
\be\psi_{l_z}'(\varphi)=\hat U_L\psi_{l_z}(\varphi)=
(2\pi a)^{-1/2}\exp{i\Bigl(l_z\varphi+\frac{\Phi}{2\Phi_0}\sin(2\varphi)\Bigr)},\label{eqpsiprime}
\ee 
with 
\be \hat U_L=\exp\Bigl({i\frac{e}{\hbar}\alpha(a,\varphi)}\Bigr)=\exp\Bigl({i\frac{\Phi}{2\Phi_0}\sin(2\varphi)}\Bigr),\ee
where the subscript $L$ is for Landau. These eigenfunctions also exhibit the ring periodicity  {(see Fig.~\ref{fig1})}.
The eigenvalues of the gauge-transformed Hamiltonian,
\be 
\hat H'=\hat U_L\hat H\hat U_L^{\dagger}=\frac{1}{2m}(-i\hbar a^{-1}\partial_\varphi-eA'_\varphi(a))^2,
\ee are of course unchanged by the unitary gauge transformation. 

In the Landau gauge, the canonical angular momentum is not the unitary transform of $\hat L_z$ (that is, $\hat L_z'=\hat L_z\ne \hat U\hat L_z\hat U^\dagger$). As a consequence, the eigenfunctions of $\hat H'$ 
are \textit{not} eigenstates of the gauge-invariant canonical angular momentum operator $\hat L_z$, 
\be
\hat L_z\psi_{l_z}'(\varphi)=\hbar\Bigl(l_z+\frac{\Phi}{\Phi_0}\cos(2\varphi)\Bigr)\psi_{l_z}'(\varphi)
\not={\tt scalar}\times\psi_{l_z}'(\varphi).\label{eqNotLz}
\ee
This result is the quantum mechanical counterpart of Eq.~(\ref{eq-lzprime}). It might be a priori surprising: we are looking at the same problem as in the unprimed gauge, so we expect the same angular momentum in a physical state of the same energy. This is indeed true, as can be observed by the calculation of the \textit{expectation value} given by the matrix element
\bey
\langle\psi_{l_z}'|\hat L_z|\psi_{l_z}'\rangle&=&\int_0^{2\pi}\!a\,d\varphi\, {\psi_{l_z}^{\prime*}}(\varphi)
\hbar\Bigl(l_z+\frac{\Phi}{\Phi_0}\cos(2\varphi)\Bigr)\psi_{l_z}'(\varphi)\nonumber\\
&=&\hbar l_z.
\eey
What Eq.~(\ref{eqNotLz}) expresses is the fact that the eigenfunctions (\ref{eqpsiprime}) are not eigenstates of the canonical angular momentum because the latter does not commute with the Hamiltonian in the Landau gauge: 
$[\hat H',\hat L_z]\not=0$.  The two operators thus cannot share the same eigenstates.

The calculation of the current density also illustrates the differences with the cylindrical gauge, although $j_\varphi=-\partial E_{l_z}/\partial\Phi$ is the same in both gauges. We have for that purpose to evaluate
\bey
j_\varphi&=&\frac em {\psi_{l_z}'}^*(\varphi)\hat \Lambda'_z\psi_{l_z}'(\varphi)\nonumber\\
&=&\frac em {\psi_{l_z}'}^*(\varphi)(-i\hbar a^{-1}\partial_\varphi-e{A'}_\varphi(a))\psi_{l_z}'(\varphi)\nonumber\\
&=&\frac{e\hbar}{2\pi ma^2}(l_z-\Phi/\Phi_0),
\eey
and the additional term due to the change of gauge $A_\varphi\to {A'}_\varphi$ is exactly compensated by the
action of $ -i\hbar\partial_\varphi$ on the modified wave function, to keep the current the same as in the cylindrical gauge.

\subsection{The case of singular gauge transformations}\label{sec3}

\begin{figure}[ht]
\begin{center}
\includegraphics[width=6.5cm]{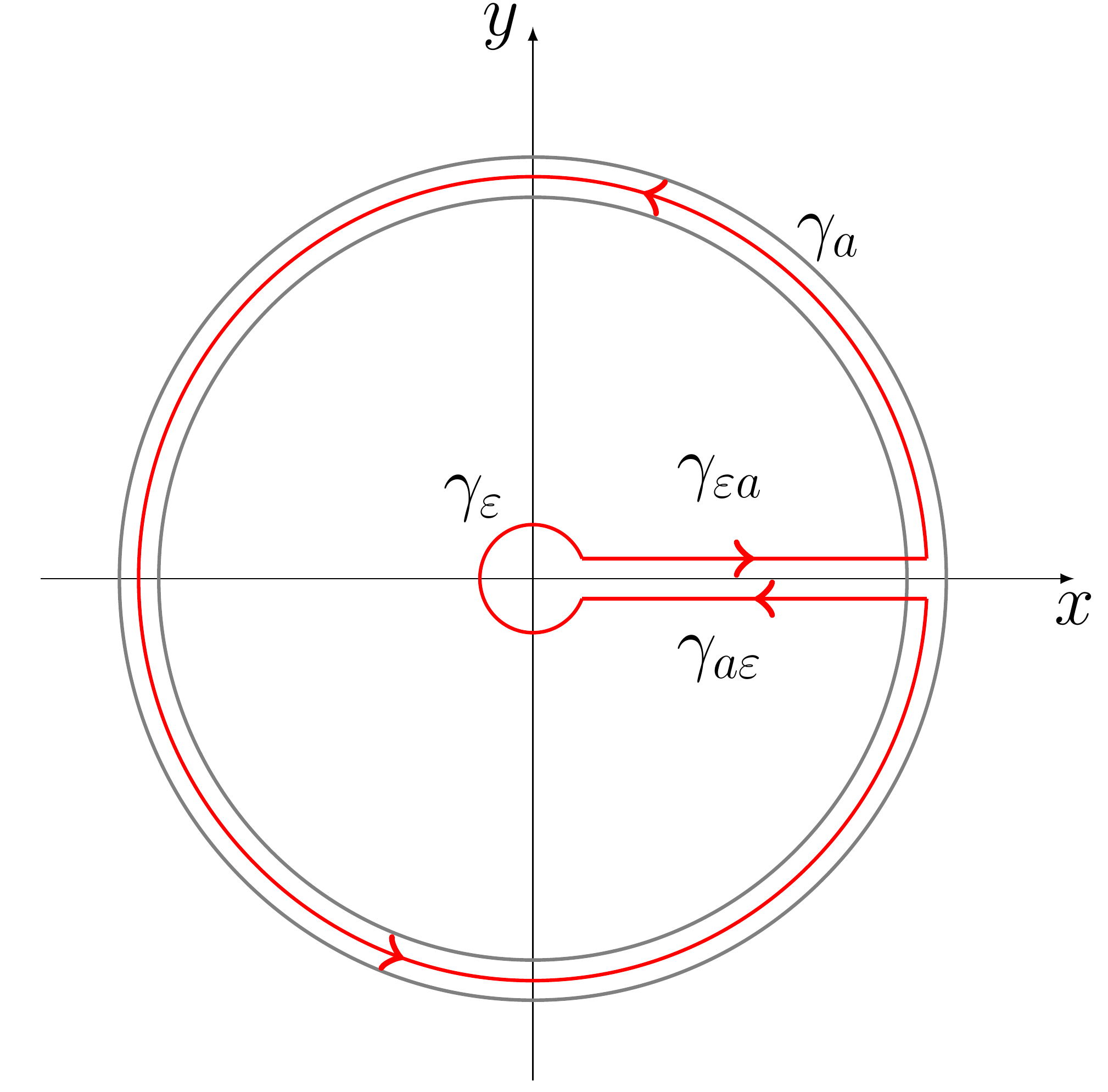}
\caption{Integration contour for the calculation of the magnetic field flux.} 
\label{figContour}
\end{center}
\end{figure}

Let us now consider a multivalued gauge transformation described by the gauge function 
\be\alpha(\varphi)=-\Phi\frac {\varphi}{2\pi}.\ee
This transformation is singular in the sense that it does not display the angular periodicity: $\alpha(\varphi+2\pi)\not=\alpha(\varphi)$.
Hence the associated unitary transformation
is also singular:
\be 
\hat U_s=e^{-i(\Phi/\Phi_0)\varphi},
\ee 
where the subscript $s$ stands for singular.
This transformation changes the gauge vector on the ring from 
$A_\varphi=\frac 12Ba$ to $A''_\varphi=0$, i.e., it appears to completely gauge-away the magnetic field, because $\int_0^{2\pi}a\,d\varphi\ \!A''_\varphi(a)=0$.  This is nevertheless not true, because the vector potential acquires a radial component $A''_\rho(\rho,\varphi)=-B\rho\varphi$ that is multivalued, and in order to close the integration circuit properly to evaluate $\int_{\cal C}\vac A''\cdot d\vac r$, one has to go around the branch cut of $A''_\rho$ (see Fig.~\ref{figContour}) and this path contributes to the circulation exactly what is needed to recover $\int_{\Sigma({\cal C})}\vac B\cdot d\vac S=\Phi$. The magnetic field is thus absent from the expression for the Hamiltonian on the ring, but still present in the wave function, as we will see.

\begin{figure}[ht]
\begin{center}
\includegraphics[width=8cm]{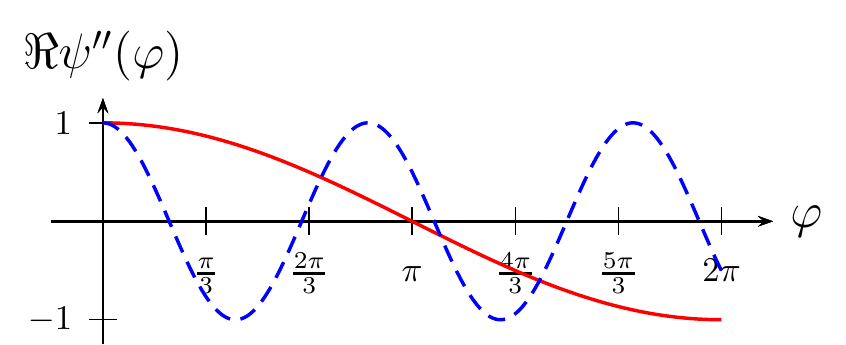}
\end{center}
\caption{The real part of the singular-gauge 
$l_z=2$ eigenfunction, for $\Phi/\Phi_0=-2.5$ (solid) and $1/3$
 (dashed). 
 }\label{fig3}
\end{figure}

Indeed, under the same transformation, the eigenfunctions become
\be \psi_{l_z}''(\varphi)=e^{-i(\Phi/\Phi_0)\varphi}\psi(\varphi)=(2\pi a)^{-1/2}e^{i(l_z-\Phi/\Phi_0)\varphi},\label{eqpsiprimeprime}\ee hence they are multivalued in the general case,\cite{Dirac31} since there is no need for the flux $\Phi$ to be equal to an integer number of flux quanta $\Phi_0$.  In the new gauge, the eigenstates belong to the Hilbert space (see Fig.~\ref{fig3})
\bey
{\cal H}''&=&\{\psi(\varphi) 
\ | \ 
 \textstyle\int_0^{2\pi}|\psi|^2 d\varphi<+\infty \nonumber\\ &&\qquad \hbox{and}\  \psi(\varphi+2\pi)=e^{-i 2\pi{\Phi}/{\Phi_0}}\psi(\varphi)\},\label{eqHilbertprimeprime} 
\eey 
which differs from Eq.~(\ref{eqHilbert}) in the boundary conditions imposed on the allowed quantum states.

The new Hamiltonian  is obtained via unitary transformation, as in the case of the non-singular gauge transformation:
\be \hat H''=\hat U_s\hat H\hat U_s^{\dagger}=\frac{1}{2m}(-i\hbar a^{-1}\partial_\varphi)^2.\label{eqHprimeprime}\ee
As claimed above, the gauge vector no longer appears in the expression for the Hamiltonian, but the magnetic flux still enters the problem via the boundary conditions and the multi-valued character of the eigenstates. 
The eigenvalues  are unchanged, 
\bey
\hat H''\psi_{l_z}''(\varphi)&=&\frac{\hbar^2}{2ma^2}(l_z-\Phi/\Phi_0)^2\psi_{l_z}''(\varphi),
\eey
and the current density, defined via
\bey
j_\varphi&=&\frac ema{\psi_{l_z}''}^*(\varphi)(-i\hbar\partial_\varphi)\psi_{l_z}''(\varphi)
\nonumber\\
&=&\frac{e\hbar}{2\pi ma^2}(l_z-\Phi/\Phi_0),
\eey
also remains unchanged.
{\textcolor{black}{
In this expression}}
 $\hbar(l_z-\Phi/\Phi_0)$ appears as the non-integer eigenvalues of the mechanical angular momentum.~\cite{Wilczek,KobePRL}

In this singular gauged problem, on the other hand, the representation of the canonical angular momentum has to be modified. Indeed, $-i\hbar\partial_\varphi$ acting on Eq.~(\ref{eqpsiprimeprime}) would not produce the proper eigenvalues $\hbar l_z$.  For the reciprocal statement, see, e.g., Ref.~\onlinecite{Shankar}: 
\begin{quote}{\textcolor{black}{The angular momentum $p_\varphi\to -i\hbar\partial/\partial_\varphi$ is Hermitian, as it stands, {\em on single-valued functions:} $\psi(\rho,\varphi+2\pi)=\psi(\rho,\varphi)$.}}
\end{quote} 
 This problem has been discussed in the literature~\cite{Kretzschmar65,Riess72} and the correct representation of the canonical angular momentum that acts in the Hilbert space of multivalued square integrable complex functions has to incorporate the boundary conditions as 
\be \hat L''_z=-i\hbar\partial_\varphi+\hbar\frac{\Phi}{\Phi_0}.
\ee
The eigenvalues of ${\hat L''}_z$ are integer multiples of~$\hbar$, as we expect from the Lie algebra of orbital angular momentum.
This property is overlooked in the literature, but it can be proven showing that this angular momentum, and not just $-i\hbar\partial_\varphi$, is indeed the  generator of rotations. A $2\pi$ rotation acting on the multivalued gauged wave functions leads to
\bey
{\cal R}_{2\pi}\psi''(0)&\equiv& e^{-\frac i\hbar 2\pi \hat L''_z}\psi''(0)\nonumber\\
&=&e^{-\frac i\hbar 2\pi (-i\hbar\partial_\varphi+\hbar\frac{\Phi}{\Phi_0})}\psi''(0)\nonumber\\
&=&e^{-i 2\pi\Phi/\Phi_0}\psi''(-2\pi)\nonumber\\
&\equiv&\psi''(0).
\eey

 \subsection{Comparison between the two approaches}
 
 The regular and singular gauge transformations are rigorously equivalent and, as expected, they correspond to two different ways of dealing with the same physical problem: either with explicit operator representation as in Eqs.~(\ref{eqH}) and (\ref{eqp}), 
 and periodic wave functions (\ref{eqpsi}) belonging to the Hilbert space (\ref{eqHilbert}); or via non-integrable phases encoded in the wave functions (\ref{eqpsiprimeprime}), which belong to the space (Eq.~(\ref{eqHilbertprimeprime})) acted on by the \textit{free-particle} Hamiltonian, 
 Eq.~(\ref{eqHprimeprime}). The first approach can be considered as the standard physicist's way of implementing a gauge interaction: Starting from the free-particle problem, the minimal coupling prescription $\hat{\vac P}=-i\hbar\bnabla\to\hat{\vac P}'=\hat{\vac P}-e\vac A=-i\hbar\bnabla-e\vac A$ is implemented in the free-particle Hamiltonian, leading to an interaction term that is apparent, and one searches for ``well behaved'' eigenfunctions, i.e., with well-defined phases. Operators there (e.g. $\hat{\vac P}=-i\hbar\bnabla$) keep their ordinary forms (they are not gauged transformed). The second approach may be considered more as following Weyl's program of geometrization of electrodynamics:~\cite{WuYang75} In this approach, the interaction is not apparent in the representation, the Hamiltonian  still being that of the free particle, but it is present in the non-integrable phase, i.e., it has been geometrized.  Dirac  has a very illuminating discussion on this question, which we highly recommend to students.~\cite{Dirac31}
 Since Dirac's exposition of the reasoning is so penetrating, we quote below a relevant excerpt:
 
 \begin{quote}
 We express $\psi$ in the form $$(2)\qquad\qquad \psi = Ae^{i\gamma}$$ where $A$ and $\gamma$ are real functions of $x,y,z$ and $t$, denoting the amplitude and phase of the wave function. For a given state of motion of the particle, $\psi$ will be determined except for an arbitrary constant numerical coefficient, which must be of modulus unity if we impose the condition that $\psi$  shall be normalised.
The indeterminacy in $\psi$ then consists in the possible addition of an arbitrary constant to the phase  $\gamma$. Thus the value of $\gamma$ at a particular point has no physical meaning and only the difference between the values of $\gamma$ at two different points is of any importance.\,\dots\  
 Let us examine the conditions necessary for this nonÐintegrability of phase not to give rise to ambiguity in the applications of the theory.\,\dots\ 
For the mathematical treatment of the question we express $\psi$ more generally than ($2$), as a
product 
 $$(3)\qquad\qquad \psi = \psi_1e^{i\beta}$$
where $\psi_1$ is any ordinary wave function (i.e., one with a definite phase at
each point) whose modulus is everywhere equal to the modulus of $\psi$. The uncertainty of phase is thus put in the factor $e^{i\beta}$.
 This requires that $\beta$ shall not be a function of $x,y,z,t$ having a definite value at each point, but $\beta$ must have definite derivatives
 $$\kappa_x=\frac{\partial\beta}{\partial x}, \ \kappa_y=\frac{\partial\beta}{\partial y}, \ \kappa_z=\frac{\partial\beta}{\partial z}, \ \kappa_0=\frac{\partial\beta}{\partial t}, \ $$
 at each point, which do not in general satisfy the conditions of integrability $\partial\kappa_x/\partial y=\partial\kappa_y/\partial x$, etc.\,\dots\ 
 From ($3$) we obtain
$$(5)\qquad\qquad -i h\frac{\partial}{\partial x}\psi=e^{i\beta}\left(-ih\frac{\partial}{\partial x}+h\kappa_x\right)\psi_1,$$
with similar relations for the $y$, $z$ and $t$ derivatives.
It follows that if $\psi$ satisfies any wave equation, involving the momentum and energy operators $\vac p$ and $W$, $\psi_1$ will satisfy the corresponding wave equation in which $\vac p$ and $W$ have been replaced by $\vac p+h\bkappa$ and $W-h\kappa_0$ respectively. 

 Let us assume that $\psi$ satisfies the usual wave equation for a free particle in the absence of any field. Then $\psi_1$ will satisfy the usual wave equation for a particle with charge $e$ moving in an electromagnetic field whose potentials are
$$(6)\qquad\qquad \vac A=\frac{hc}{e}\bkappa,\quad\vac A_0=-\frac{h}{e}\kappa_0.$$
Thus, since $\psi_1$ is just an ordinary wave function with a definite phase, our theory reverts to the usual one for the motion of an electron in an electromagnetic field. This gives a physical meaning to our non-integrability of phase. 
We see that we must have the wave function $\psi$ always satisfying the same wave equation, whether there is a field or not, and the whole effect of the field when there is one is in making the phase nonÐintegrable.
The components of the 6-vector $ \vac{curl}\ \!\bkappa$ 
are, apart from numerical coefficients, equal to the components of the electric and magnetic fields $\vac E$ and $\vac H$. They are, written in three-dimensional vector-notation,
$$(7)\qquad\qquad\vac {curl}\ \!\bkappa =\frac{e}{hc}\vac H,\quad\vac{grad}\ \!\kappa_0-\frac{\partial\bkappa}{\partial t}=\frac{e}{h}\vac E.$$

 The connection between non-integrability of phase and the electromagnetic field given in this section is not new, being essentially just Weyl's Principle of Gauge Invariance in its modern form.
 \end{quote}

\section{Discussion: Geometrization of physics}

\begin{figure}[ht]
\begin{center}
\includegraphics[width=9cm]{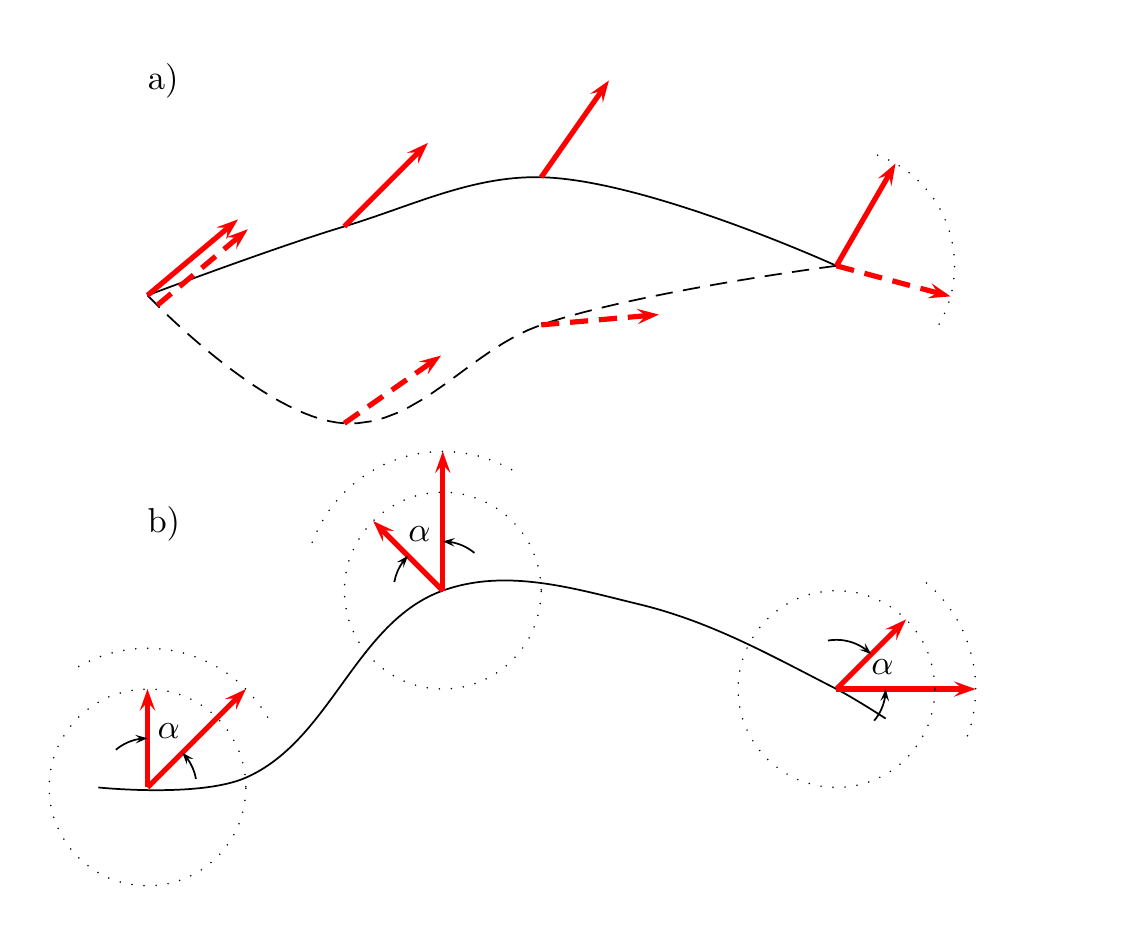}
\caption{Einstein gravitation.  (a) Parallel transport of a vector between the same starting and ending points along two distinct curves (solid and dashed) leads to vectors of different orientations in a curved space.  (b) Parallel transport keeping two vectors' lengths conserved and relative orientations fixed.} 
\label{FigGTR}
\end{center}
\end{figure}

The first theory in which gauge symmetry plays its full role as we understand it today is Einstein's theory of gravitation, general relativity.  There, the gravitational interaction is ``geometrized,'' i.e., instead of considering the motion of a point particle in space-time, subject to gravitational interaction with, for example, massive particles, the point particle follows geodesics, which are just free-fall trajectories in a curved space-time with metric $ds^2=g_{\mu\nu}dx^\mu dx^\nu$. Free fall is understood as the motion of a free particle, its Lagrangian being purely kinetic energy. The interaction enters, via the gravitational potential, into the metric tensor $g_{\mu\nu}$ of the curved space-time according to Einstein's field equations.~\cite{Einstein1915} The geometry of the underlying space-time is Riemannian geometry, in which vector lengths are invariant (e.g., $ds^2$), but their orientation (the phase in our electromagnetic examples) is not integrable, i.e., two vectors that follow parallel transport along a space-time curve will see their orientations change in a curved manifold in a manner that is path dependent (non-integrability of the orientation), but their relative orientation will remain unchanged (see Fig.~\ref{FigGTR}).

\begin{figure}[ht]
\begin{center}
\includegraphics[width=9cm]{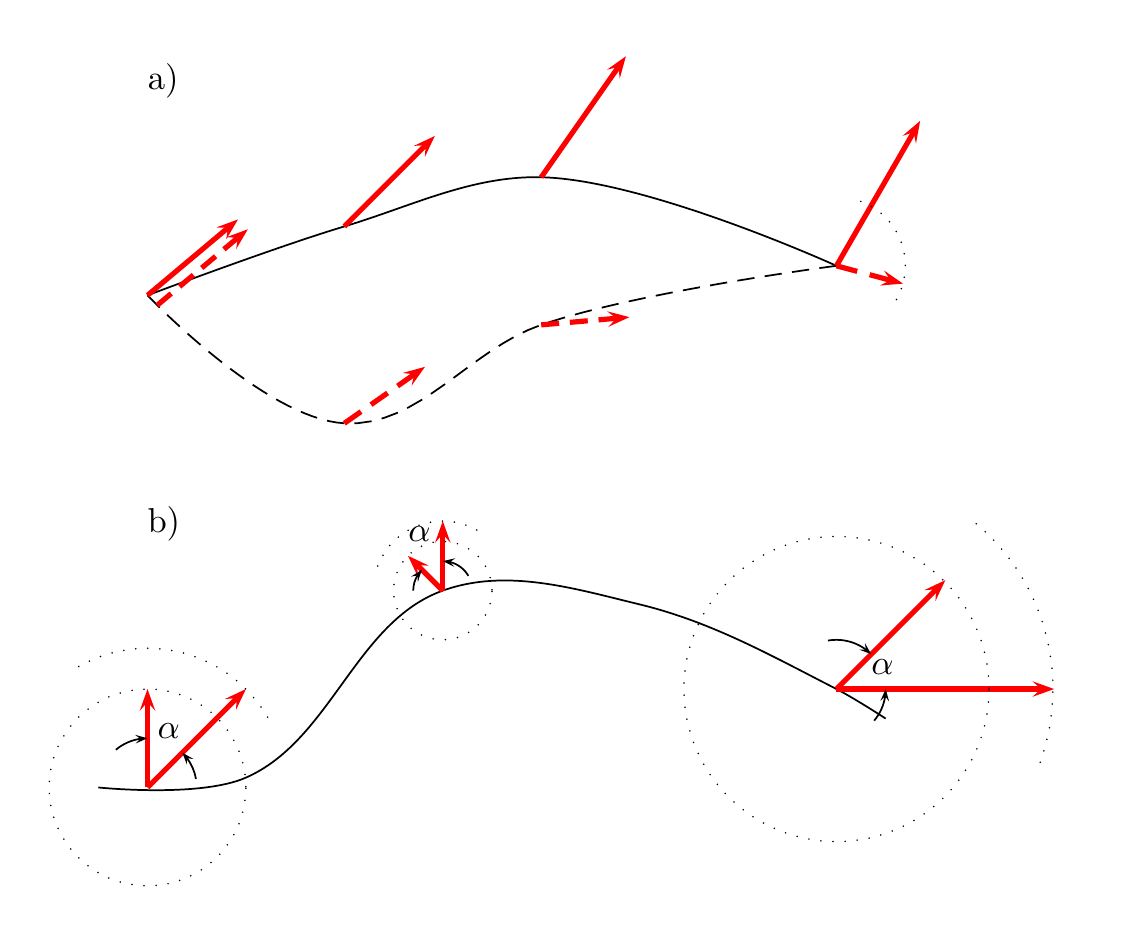}
\caption{ Weyl gravitation. (a) Parallel transport of a vector between the same starting and ending points along two distinct curves leads to vectors of different orientations \textit{and lengths}.
(b) Parallel transport, not conserving vector lengths, but keeping length ratios and relative orientations preserved.} 
\label{FigWeylG}
\end{center}
\end{figure}

The gauge principle was later elaborated by Weyl in his 1918 paper~\cite{Weyl1918} where he considered, in addition to the quadratic form $ds^2$, a linear form $d\ell = \ell_\mu dx^\mu$, which enables measuring the length of vectors, and he relaxed the constraint of length conservation of Riemann geometry. Now, not only the orientation but also the length of a vector is non-integrable: two vectors that follow parallel transport along a space-time curve will now see their lengths and their orientations change in a path-dependent manner, but their relative lengths and relative orientation remain unchanged (see Fig.~\ref{FigWeylG}). The quantity $\ell_\mu$, which allows for this non-integrability of length, was identified by Weyl as $\ell_\mu=(e/\gamma) A_\mu$, where $A_\mu$ is the gauge vector of electrodynamics, {\textcolor{black}{and $\gamma$ is an unknown constant with the dimensions of an action}}.  This identification is similar to the way $g_{\mu\nu}$ encodes the gravitational potential in Einstein theory.  The square of the length $||v||^2=g_{\mu\nu}v^\mu v^\nu$ of a vector $v^\mu$ transported between points $A$ and $B$  along a curve ${\cal C}$ is now path dependent and is determined by the ``gauge field'' $A_\mu$: $||v_B||^2=||v_A||^2 \exp\bigl((e/\gamma)\int_{\cal C}A_\mu dx^\mu\bigr)$.

In spite of its beautiful mathematical construction, incorporating gravitation and electromagnetism in a single unified theory, Weyl's approach did not survive Einstein's criticism since it was unsuccessful at describing the physical world: It predicted that the time measured by a clock (e.g., frequencies given by atomic spectra) would depend on its history, a prediction that has never been observed experimentally.

Soon after Weyl's work, Schr\"odinger, London, and Fock noticed that Weyl's theory could be adapted to quantum mechanics,~\cite{O'Raifeartaigh} essentially at the price of allowing the constant prefactor between $\ell_\mu$ and $A_\mu$ to be purely imaginary.  Instead of non-integrable lengths, the theory now turns into one with non-integrable phases, as was synthesized by Weyl in his 1929 paper,\cite{Weyl1929} which can be considered as the birth of modern gauge theory.  The mathematical object that is now transported is the wave function $\psi$, and the constant $\gamma$, as discussed by Schr\"odinger, with dimensions of an action, is identified as $\hbar$:
\be 
\psi=\psi_0e^{i(e/\hbar)\int_{\cal C}A_\mu dx^\mu}.
\ee
The connection $A_\mu$ that acts to connect  vectors or wave functions between different space-time points is peculiar in the sense that it is a non-integrable function. This means
that the function has no definite value at each point in space while it has a definite derivative. In the different contexts we have
discussed, the orientation of a vector in general relativity, the length of the vector in Weyl's theory, and the phase
of the wave function in quantum mechanics are all non-integrable in this sense.


\begin{acknowledgments} 
EM and BB are respectively grateful to the University of Lorraine and to IVIC for invitations. They also thank the CNRS and FONACIT 
for support through the ``PICS'' programme {\it Spin transport and spin manipulations in condensed matter: polarization, spin currents and entanglement}. BB and DM benefited from useful discussions with the group ``Connexions'' between mathematicians, historians, and physicists at Universit\'e de Lorraine.
\end{acknowledgments}

\section*{References} 

\def\paper#1#2#3#4#5{{#1,}{\ {\it #2}}{\ {\bf #3},}{\ #4}{\ (#5).}}
\def\papertitle#1#2#3#4#5#6{{#1,}{\ {\it #2}}{\ {\bf #3},}{\ #4}{\ (#5).}}

\end{document}